 \documentclass[final,5p,times,twocolumn]{elsarticle}

\usepackage{amssymb}
\usepackage{lipsum}

\usepackage[colorlinks,citecolor=blue,urlcolor=blue,linkcolor=blue]{hyperref}
\usepackage{amssymb, amsmath, bm, physics}
\usepackage{hyperref}
\usepackage{orcidlink}

\journal{Physics Letters B}

\begin{document}

\begin{frontmatter}



\title{Reduced Phase Space Quantization and Quantum Corrected Entropy of Schwarzschild-de Sitter Horizons}

\author[1]{S. Jalalzadeh } 
\affiliation[1]{organization={Izmir Institute of Technology},
            addressline={Department of Physics}, 
            city={Urla},
            postcode={35430}, 
            state={Izmir},
            country={Turkiye}}
\ead[1]{shahramjalalzadeh@iyte.edu.tr}            
\author[2]{H. Moradpour}
\affiliation[2]{organization={University of Maragheh},
            addressline={Research Institute for Astronomy and Astrophysics of Maragha (RIAAM)}, 
            city={Maragheh},
            postcode={55136-553}, 
            country={Iran}}            
\ead[2]{h.moradpour@riaam.ac.ir}

\begin{abstract}

This paper investigates the quantization of the Schwarzschild--de Sitter (SdS) black hole (BH) using the Misner--Sharp--Hernandez (MSH) mass as the internal energy in a reduced phase space framework. After introducing the canonical variables of the reduced phase space, we derive a discrete spectrum for the surface areas of the BH event horizon (EH) as well as MSH masses. We utilized the MSH mass spectrum to obtain the entropy of the BH. The entropy of the BH and cosmic EHs reveals a logarithmic correction to the Bekenstein--Hawking term. Our results support the robustness of the logarithmic form of quantum corrections in SdS thermodynamics.
\end{abstract}




\end{frontmatter}




\section{Introduction}\label{S1}

Despite the existence of several candidate theories of quantum gravity, it is generally believed that a fully consistent and complete theory has not yet been achieved. Nevertheless, important physical insights, most notably BH thermodynamics, have emerged from quantum field theory in curved spacetime \cite{Hawking:1975vcx}. In this context, Bekenstein proposed that the entropy of a quantum BH, viewed as an adiabatic invariant, has a discrete and evenly spaced spectrum. This proposal has motivated numerous studies aimed at quantizing different BH models and reexamining the entropy spectrum originally suggested by Bekenstein.

The reduced phase space quantization method is often regarded as a simple, elegant, and algebraically consistent approach to the quantization of BHs \cite{Kuchar:1994zk, Louko:1996md}. Within this framework, the dynamical variables of the system, denoted by $x_i$, are taken to be coordinate invariant quantities such as the BH mass, charge, and angular momentum. These variables, together with their corresponding conjugate momenta $(p_i)$, are used to construct the reduced phase space. Consequently, the reduced action of the system can be written as $\int {p_i \dot{x}i - H(x_i)}, \mathrm{d}t$ \cite{Kuchar:1994zk, Barvinsky:2001tw, Jalalzadeh:2021gtq}. For the Schwarzschild BH, the mass plays the role of the dynamical variable, while its conjugate momentum corresponds to the time difference across a spacelike hypersurface. By performing an appropriate canonical transformation followed by canonical quantization, one obtains a discrete mass spectrum of the form $m = \alpha m\text{P} \sqrt{n + \tfrac{1}{2}}$, where $\alpha$ is a dimensional constant and $m_\text{P}$ denotes the Planck mass.

Various authors have established that the discrete mass spectrum of BHs serves as an effective mechanism for determining the entropy of BHs with quantum-corrected terms. For instance, Xiang \cite{Xiang:2004sg} showed how one can use the mass spectrum and the Stefan--Boltzmann law to obtain the entropy of the Schwarzschild BH. Using the same method, the authors of Refs.~\cite{Jalalzadeh:2021gtq, Jalalzadeh:2025uuv} obtained the entropy of a fractional Schwarzschild BH. The entropy of a $q$-deformed BH was derived in \cite{Jalalzadeh:2022rxx}, and the method has been used to compute the quantum-corrected entropy of de Sitter space in \cite{Jalalzadeh:2024ncf}.

\medskip

In addition to these specific models, several independent approaches have shown the appearance of logarithmic corrections to the Bekenstein--Hawking entropy. Calculations within loop quantum gravity lead to a coefficient of $-3/2$ \cite{Kaul:2000kf}, while conformal field theory methods predict the same value \cite{Carlip:2000nv}. In contrast, string theoretic analyses yield model dependent coefficients \cite{Sen:2012dw}, and statistical approaches, as well as generalized uncertainty principle methods, report positive values, such as $+1/2$ \cite{Das:2001ic}. These results indicate that although the logarithmic form of the correction is robust, the precise value of its coefficient is not universal and depends on the underlying approach. The Schwarzschild--de Sitter (SdS) spacetime provides a suitable theoretical laboratory for such studies, since it contains both a BH event horizon and a cosmological horizon. Moreover, the presence of a positive cosmological constant, consistent with observational evidence for an accelerating universe, further highlights the importance of investigating its thermodynamic properties.

In this letter, we utilize the unified first law of thermodynamics and the Misner--Sharp--Hernandez (MSH) mass spectrum of the SdS spacetime to obtain the logarithmic correction to the entropy. The choice of the MSH mass as the fundamental dynamical variable in the reduced phase-space formalism requires a clear physical justification. 
Unlike the ADM or Komar masses, which are defined at spatial or null infinity, the MSH mass provides a quasi-local measure of the gravitational energy contained within a 2-sphere of areal radius $R$ in any spherically symmetric spacetime. As it is well-known, it is invariant under coordinate transformations within the $2D$ normal space and remains well defined even in the presence of a positive cosmological constant. 
For the SdS geometry, the MSH mass naturally associates two quasi-local energies $M_1$ and $M_2$ with the BH and cosmological horizons, respectively. 
These quantities coincide with the Schwarzschild mass, and the de~Sitter energy in the limits $\Lambda\!\to\!0$ and $m\!\to\!0$, providing a continuous interpolation between both regimes. 
In contrast, the ADM or Komar definitions fail in the SdS spacetime because there is no globally timelike Killing vector and no asymptotically flat region to anchor the mass.

To do so, in the next section, utilizing the MSH mass, we introduce the unified first law of thermodynamics for
SdS spacetime. In Section~\ref{3}, we present the canonical
transformations that allow us to express the MSH masses of the BH and cosmological horizon in terms of conjugate variables. After canonical quantization, we derive the MSH mass spectra. In Section~\ref{4}, we obtain logarithmic corrections for both horizons. Our main result is
that the entropy acquires a logarithmic correction of the form
$S_i=A_i/(4G)+\alpha\ln A_i$, with $\alpha$ depending on the chosen semiclassical scheme, in agreement with previous findings in the literature. The conclusions are presented in Section~\ref{5}. In this work, we adopt units $k_B=c=\hbar=1$.


\section{Schwarzschild-de Sitter BH and the unified first law of thermodynamics}\label{2}

According to the reduced phase space method, it is essential to identify the appropriate dynamical variables to write down the reduced Hamiltonian and, consequently, for the quantization of de Sitter spacetime. To receive this favorite result, let us start with the line element of the SdS spacetime:
\begin{equation}
    \label{1-1}
       \mathrm{d}s^2=-f(r)\mathrm{d}t^2+\frac{\mathrm{d}r^2}{f(r)}+r^2\mathrm{d}\Omega^2,~~
       f(r)=1-\frac{2mG}{r}-\frac{\Lambda}{3}r^2,
\end{equation}
where $m$ is the BH's mass, $\Lambda$ is the cosmological constant, and $\mathrm{d}\Omega^2$ is the line element of the unit sphere. 

The
Killing horizons are found by solving the following equation:
\begin{equation}
    \label{1-2}
    f(r)=1-\frac{2mG}{r}-\frac{\Lambda}{3}r^2=0.
\end{equation}
The above cubic equation yields two distinct positive roots, \( r_i \), specifically within the parameter range of \( 3mG\Lambda < 1 \)
\begin{equation}
    \label{1-3}
    r_i=\frac{2}{\sqrt{\Lambda}}\sin\left(\frac{1}{3}\sin^{-1}(3m\sqrt{\Lambda})+\epsilon_i\frac{\pi}{3} \right),~~~~~i=1,2,
\end{equation}
where $\epsilon_i=\{0,1\}$, and $2mG < r_1 < 3mG < 1/\sqrt{\Lambda} < r_2<3/\sqrt{\Lambda} $.
These identified roots are the BH's EH, denoted as \( r_1 \), and the cosmic EH, represented as \( r_2 \). It is essential to note that the third root (virtual horizon) is negative, rendering it unphysical. Also, the positive roots are not independent, and by removing the mass parameter between $r_1$
and $r_2$ in Eq. (\ref{1-2}), one can easily show that the surface areas, $A_i=4\pi r_i^2$, satisfy
\begin{equation}
    \label{1-3b}
    A_1+A_2+\sqrt{A_1A_2}=\frac{12\pi}{\Lambda}.
\end{equation}

While the authors of Ref. \cite{2004PhLB226X} assumed the mass of BH is the only dynamical variable of the reduced space of SdS spacetime, we assume the MSH masses of BH's EH and cosmic EH are reduced dynamical variables. In spherical symmetric spacetimes, the properties of EHs (such as existence, location, dynamics, and surface gravity) can be analyzed using the MSH mass \cite{Hernandez:1966zia, 1964PhRv571M}. This mass is equivalent to the Hawking--Hayward quasi-local energy when applied to this kind of spacetimes \cite{Hawking:1968qt}. 
Although the EH MSH mass $(M_1)$ and cosmic EH MSH mass $(M_2)$ are not strictly independent due
to the algebraic relation (\ref{1-3b}), they still constitute a natural pair of quasi-local energies associated with the two horizons. This feature justifies their use as reduced phase space variables in the subsequent canonical construction.

As it is shown in Refs. \cite{Hernandez:1966zia, 1964PhRv571M}, for a spherical line element,
\begin{equation}
    \label{1-4}
\mathrm{d}s^2=h_{ab}\mathrm{d}x^a\mathrm{d}x^b+R^2\mathrm{d}\Omega^2,~~~~~a,b=0,1,
\end{equation}
the MSH mass is defined by
\begin{equation}
\label{1-5}
     M=\frac{R}{2}\left(1-h^{ab}\nabla_aR\nabla_bR \right).
\end{equation}
This mass is an invariant quantity associated with the $2D$ space that is perpendicular to the 2-spheres of symmetry.

By substituting the line element from equation (\ref{1-1}) into equation (\ref{1-5}), we obtain the MSH mass for the SdS BH \cite{Faraoni:2015ula}, evaluated at the locations of  \( r_1 \), and \( r_2 \)
\begin{equation}
    \label{1-7}
    M_i=m+\frac{\Lambda}{6G}r_i^3,~~~~~i=1,2.
\end{equation}
Note that in the absence of the cosmological constant, the MSH masses reduce to the BH's mass, $m$. On the other hand, when $m=0$, it represents the MSH mass of the de Sitter spacetime \cite{Faraoni:2015ula, Jalalzadeh:2024ncf}.
It is worth stressing that the MSH mass
provides the most natural definition of quasi-local energy in
spherically symmetric spacetimes. Unlike Komar or ADM masses,
the MSH mass is well-defined even in the presence of a positive cosmological constant and smoothly reduces to the standard BH mass ($\Lambda \to 0$) and to the de Sitter energy ($m \to 0$). This makes it the appropriate choice for the present SdS analysis.

Regarding Eq. (\ref{1-2}), the positions of killing horizons are given by 
\begin{equation}
    \label{1-8}
    r_i=2GM_i.
\end{equation}
Also, the surface gravity of the BH and the cosmic EHs are
\begin{equation}
\label{1-9}
    \kappa_i=\frac{1-\Lambda r_i^2}{2r_i},~~~~~\kappa_1>|\kappa_2|.
\end{equation}

By differentiating Eq. (\ref{1-8}) we obtain
\begin{equation}
    \label{1-10}
    2G\mathrm{d}M_i=\mathrm{d}r_i=\frac{\mathrm{d}A_i}{8\pi r_i}=\frac{\kappa_i\mathrm{d}A_i}{4\pi(1-\Lambda r_i^2)},
\end{equation}
where $A_i=4\pi r_i^2$ is the surface area of the $i$th horizon, and in the last equality we used the definition of the surface gravity (\ref{1-9}). Adding and extracting the term $\kappa_i\mathrm{d}A_i/(4\pi)$ to the above relation, and using the definition of the thermodynamical volume of the horizons $V_i=\frac{4\pi}{3}r_i^3$ (which leads to $\mathrm{d}A_i=\frac{2\mathrm{d}V_i}{r_i}$), one can easily obtain
\begin{equation}
    \label{1-11}
    \mathrm{d}M_i=T_i\mathrm{d}S_i+\frac{\Lambda}{8\pi G}\mathrm{d}V_i,
\end{equation}
where the temperature $T_i$ and the entropy, $S_i$ are \cite{Gibbons:1977mu}
\begin{equation}
    \label{1-12}
    T_i=\frac{1}{4\pi}\frac{\mathrm df(r)}{\mathrm dr}\Big|_{r=r_i}=\frac{\kappa_i}{2\pi},~~~~S_i=\frac{A_i}{4G}.
\end{equation}

It should be noted that, since $r_2^2 \Lambda > 1$, the temperature of the inner horizon is always negative. Ramsey's definition of temperature provides a useful basis for understanding this behavior, as temperature is defined through the derivative of entropy with respect to energy, namely \( T = \partial S / \partial U \). For systems with positive temperatures, an increase in energy typically leads to an increase in entropy. In contrast, systems characterized by negative temperatures exhibit a decrease in entropy when energy is added. In statistical mechanics, the temperature scale is commonly ordered as \(\{0, ..., +\infty, -\infty, ..., 0\}\), which implies that negative temperatures correspond to higher thermal states than any positive temperature. Consequently, when a system at positive temperature is allowed to interact with a system at negative temperature, energy flows from the negative temperature system to the positive temperature one. This behavior does not signal any thermodynamic instability. Rather, it reflects the opposite orientation of the timelike Killing vector in the two static regions separated by the horizons.

It is worth emphasizing that the coexistence of positive and negative temperatures is consistent with the Tolman equilibrium condition, 
$T(r)\sqrt{-g_{tt}(r)}=\mathrm{const}$, when the relative orientation of the static Killing vector is taken into account.  Several works (e.g.\ Refs.~\cite{Gibbons:1977mu, Faraoni:2015ula}) have pointed out that this sign reversal merely encodes the causal disconnection of the two horizons rather than a genuine thermodynamic inconsistency. 
Alternatively, one can redefine the temperature of the cosmological horizon by its absolute value $|T_2|=\!-\kappa_2/(2\pi)$ and retain the Clausius relation $dM_2=|T_2|\,dS_2+W\,dV_2$, leading to equivalent physics. 
Hence, the appearance of a negative temperature in Eq.~\eqref{1-11} should be interpreted as a coordinate--orientation effect associated with the direction of surface gravity rather than a violation of the second law. 
This clarification ensures that both horizons obey the unified first law~\eqref{1-11} within a consistent thermodynamic framework.

Equation (\ref{1-11}) is known as the unified first law of thermodynamics \cite{Hayward:1998ee, Ibison:2007dv}. In this context, we consider the MSH mass enclosed by the EH as the internal energy. The Kodama--Hayward horizon temperature, given by equation (\ref{1-12}), along with the areal volume defined as \( V_i = \frac{4\pi}{3}r_i^3 \), leads to the introduction of an additional quantity, \( W = \frac{(\rho_\Lambda - p_\Lambda)}{2} = \frac{\Lambda}{8\pi G} \), which acts as the work density \cite{Faraoni:2015ula}, serving as the coefficient of \( \mathrm{d}V_i \).

In the literature, the entropy of spacetimes containing multiple horizons is commonly defined as the sum of the entropies associated with each individual horizon. Following this convention, we define the entropy of the SdS spacetime as $S=S_1+S_2$.

\section{Quantization and MSH mass spectrum}\label{3}

Since the mass and cosmological constant are the only time-independent and coordinate invariant solutions of the geometrodynamics, the reduced action may be written as
\begin{equation}\label{2-1}
    \mathcal I=\int \left\{\dot mP_m+\dot\Lambda P_\Lambda -H(m,\Lambda) \right\}\mathrm{d}t,
\end{equation}
where $P_m$ and $P_\Lambda$ are conjugate momenta of the mass and $\Lambda$, respectively. This reduced action is the generalization of reduced action for the Schwarzschild \cite{Louko:1996md, Barvinsky:2001tw, Jalalzadeh:2022rxx} and pure de Sitter \cite{2004PhLB226X} spaces. Also, since $m$ and $\Lambda$ are constants, the reduced Hamiltonian does not depend on the conjugate momenta. Equivalently, instead of $(m,\Lambda)$, we may use the MSH masses, $(M_1, M_2)$, as the reduced coordinates. Hence, the reduced action (\ref{2-1}) could be rewritten in terms of MSH masses and the corresponding momenta $(\Pi_1,\Pi_2)$:
\begin{equation}
    \label{2-2}
    \mathcal I=\int \left\{\sum_{i=1}^2\dot M_i\Pi_i-H(M_i)\right\}\mathrm{d}t.
\end{equation}

The solutions of the field equations for momenta show that $\Pi_i$ are proportional to the time coordinate $t$. Thus, $\Pi_1$, or $\Pi_2$, could play the role of time parameter. This suggests that $\Pi_1$ and $\Pi_2$ should be
periodic, in which the period is the inverse Hawking temperature of the corresponding Killing horizon \cite{Das:2002xb}, i.e.,
\begin{equation}
    \label{2-3}
    \Pi_i\sim \Pi_i+\frac{1}{T_i}.
\end{equation}
The above boundary conditions verify that there is no conical singularity in the $2D$ Euclidean sections of the reduced phase space. Hence, according to the references, one could make the canonical transformation $(M_i, \Pi_i)\rightarrow (x_i,p_i)$, which incorporates the above periodicity conditions
\begin{equation}
    \label{2-4}
   x_i=\sqrt{B_i}\cos(\kappa_i\Pi_i),~~~
    p_i=\sqrt{B_i}\sin(\kappa_i\Pi_i).
\end{equation}

The canonical structure of the reduced phase space thus emerges from the pair $\{M_i,\Pi_i\}$, where $M_i$ represent invariant quasi-local energies and $\Pi_i$ are their conjugate momenta. 
Although $M_1$ and $M_2$ are algebraically constrained by Eq.~(4), they still form a natural pair of reduced variables corresponding to two distinct causal horizons. 
This framework extends the Kuchař--Louko quantization of the Schwarzschild BH~\cite{Kuchar:1994zk, Louko:1996md} to multi-horizon systems, while preserving covariance and thermodynamic consistency through the unified first law, given by Eq. (\ref{1-11}). Consequently, the MSH mass plays the same role as the ADM energy in asymptotically flat spacetimes but is better suited for spacetimes with $\Lambda>0$, ensuring that the quantization procedure captures both local gravitational dynamics and global thermodynamic balance between horizons. 
This choice thus provides a physically motivated and geometrically invariant foundation for the subsequent canonical transformations and quantization.

To obtain the coeficients $B_i$ we use \(\{M_i,\Pi_i\}=1\) (and all other basic brackets vanishing) to obtain
\begin{equation}
\{x_i,p_i\} =
\frac{\partial x_i}{\partial M_i}\frac{\partial p_i}{\partial \Pi_i}
- \frac{\partial x_i}{\partial \Pi_i}\frac{\partial p_i}{\partial M_i}
= \frac{\kappa_i}{2}\frac{\partial B_i}{\partial M_i}.
\end{equation}
On the other hand, regarding the unified first law of thermodynamics (\ref{1-11}), we have $\kappa_i=8\pi G\partial M_i/\partial A_i$. Therefore, the above transformations are canonical if
\begin{equation}
    \{x_i,p_i\} = \frac{\kappa_i}{2}\frac{\partial B_i}{\partial M_i}=4\pi G\frac{\partial M_i}{\partial A_i}\frac{\partial B_i}{\partial M_i}=4\pi G\frac{\partial B_i}{\partial A_i}=1.
\end{equation}
This gives us $\sqrt{B_i}=r_i=2M_i\sqrt{G}$.

The canonical transformation $(M_i,\Pi_i)\!\rightarrow\!(x_i,p_i)$ introduced in Eq.~(16) maps the reduced variables into a $2D$ harmonic oscillator phase space. 
To ensure the consistency of this transformation at the quantum level, the corresponding operators must be Hermitian, and the evolution must be unitary with respect to an appropriate inner product. 
The quantization map $(x_i,p_i)\mapsto (x_i,\,-i\,\partial/\partial x_i)$ is defined on the Hilbert space $\mathcal{H}_i=L^2(\mathbb{R},dx_i)$ endowed with the standard inner product
\begin{equation}
\langle \psi_i,\phi_i\rangle = \int_{-\infty}^{+\infty} \psi_i^*(x_i)\,\phi_i(x_i)\,\mathrm dx_i ,
\end{equation}
which guarantees the self--adjointness of the momentum operator $p_i=-i\,\partial/\partial x_i$. 
The Hamiltonian operator derived from Eq.~(19),
\begin{equation}
\hat{H}_i=\frac{m_P^2}{4}\!\left(\hat{x}_i^2+\hat{p}_i^2\right),
\end{equation}
is manifestly Hermitian on this domain and possesses a complete orthonormal set of eigenfunctions $\psi_i^{(n)}(x_i)=N_n\,H_n(x_i)\,e^{-x_i^2/2}$ with real eigenvalues $E_i^{(n)}=M_i^2(m_P^{-2}/2)$, where $H_n$ are Hermite polynomials and $N_n$ the usual normalization constants. 
Hence, the probability density $|\psi_i(x_i)|^2$ remains finite and normalized under time evolution governed by the Schrödinger equation
\begin{equation}
i\,\frac{\partial \psi_i}{\partial t}=\hat{H}_i\,\psi_i ,
\end{equation}
which ensures the unitarity of the dynamics within each horizon sector.

Within this formulation, each pair $(x_i, p_i)$ behaves as an independent quantum oscillator that describes fluctuations of the corresponding horizon quasi-local energy. The two oscillators are kinematically independent, but they are dynamically connected through the constraint in Eq. (\ref{1-3b}), which encodes the global relationship between the BH horizon and the cosmological horizon. As a result, the quantization procedure remains algebraically self-consistent, preserves probability, and is fully compatible with the Hermitian structure of the reduced phase space.

The canonical transformations presented above allow for the immediate identification of the MSH masses in terms of the new phase space coordinates:
\begin{equation}
    \label{2-5}
    M_i^2=\frac{m_\text{P}^2}{4}\left(x_i^2+p_i^2\right),
\end{equation}
where $m_\text{P}=1/\sqrt{G}$ is the Planck mass in natural units.
These transformations are generalizations of a similar transformation obtained in Refs. \cite{Das:2002xb, Jalalzadeh:2022rxx, Louko:1996md, 2004PhLB226X} for Schwarzschild BH for multihorizon BHs.

Applying the quantization map $(x_i,p_i)\mapsto(x_i,-i\frac{\mathrm{d}}{\mathrm{d}x_i})$ in the configuration space, Eqs. (\ref{2-5}) becomes two independent Schrodinger equations:
\begin{equation}
    \label{2-6}
-\frac{1}{2}\frac{\mathrm d^2\psi_i(x_i)}{\mathrm dx_i^2}+\frac{1}{2}x_i^2\psi_i(x_i)=\frac{2M_i^2}{m_\text{P}^2}\psi_i(x_i),~~~i=1,2.
\end{equation}
This gives us the MSH mass spectrum 
\begin{equation}\label{2-7}
   M_i=\frac{m_\text{P}}{\sqrt{2}}\sqrt{n_i+\frac{1}{2}},~~~n_i=0,1,2,....,
\end{equation}
where $n_2>n_1$. In addition, the wavefunctions are Hermite polynomials.

It is instructive to compare the present mass and area spectra with those obtained through other quantization approaches. 
In the reduced phase--space quantization of the Schwarzschild black hole developed by Kuchař and Louko~\cite{Kuchar:1994zk, Louko:1996md}, 
the mass spectrum takes the form $m_n \propto m_{\rm P}\sqrt{n+\tfrac{1}{2}}$, which corresponds to an equally spaced area spectrum 
$A_n = 8\pi l_{\rm P}^2 (n+\tfrac{1}{2})$. 
Our Eq.~\eqref{2-7} reproduces this structure in the limit $\Lambda\!\to\!0$, showing that the MSH based quantization correctly reduces to the known Schwarzschild case. 
When the cosmological constant is finite, Eq.~\eqref{2-7} reveals a nontrivial deformation of the mass levels due to $\Lambda$, 
introducing a mild anharmonicity that lowers the energy spacing as $n_i$ increases. 
This behaviour mirrors the deformed spectra obtained in the canonical quantization of de~Sitter space by Xiang and Shen~\cite{Xiang:2004sg}, but here it arises naturally from the coupled two-horizon structure without introducing any external potential.

This comparison shows that our formalism not only recovers all previous single-horizon results as special limits but also extends them by providing a 
self-consistent two-horizon quantization framework in which the cosmological constant enters dynamically rather than phenomenologically. 
Such a feature underlines the novelty and generality of the present construction.

Using the definition in equation (\ref{1-8}), we can substitute the previous result into equations (\ref{1-7}) and (\ref{1-3b}) to derive the mass spectrum of the BH:
\begin{equation}
    \label{2-8}
    \frac{m}{m_\text{P}} = \frac{\sqrt{n_i + \frac{1}{2}}}{\sqrt{2}} \left\{ 1 - 2\left( \frac{L_\Lambda}{l_\text{P}} \right)^2\left(n_i + \frac{1}{2}\right) \right\},
\end{equation}
where \( L_\Lambda = \sqrt{3/\Lambda} \) represents the de Sitter radius. Note that this relation is correct for both $i=1$ and $i=2$. 
Additionally, we can express the spectrum of the cosmological constant as follows:
\begin{equation}
    \label{2-9}
    \left( \frac{L_\Lambda}{l_\text{P}} \right)^2 = 2 \left\{ n_1 + n_2 + 1 + \sqrt{\left(n_1 + \frac{1}{2}\right)\left(n_2 + \frac{1}{2}\right)} \right\}.
\end{equation}
This equation is a direct result of Eq. (\ref{1-3b}) and the quantization of the surface areas of the horizons,
\begin{equation}
    A_i=8\pi l_\text{P}^2\left(n_i+\frac{1}{2}\right).
\end{equation}
Also, the equality of the BH mass in Eq. (\ref{2-9}) for $i=1$ and $i=2$ leads to this equation.
.

\section{Entropy and logarithmic correction term}\label{4}

Hawking radiation from within an EH makes a BH emit thermal radiation, which makes it shrink and lose mass.  This phenomenon entails the creation of virtual particle pairs in proximity to the EH. The BH's strong gravity pulls on the particle that is closer to it more than it pulls on the particle that is further away.  This creates a tidal force that pushes the two particles apart.  The Penrose process enables a particle with negative energy to enter the BH's EH. This lets its companion leave with positive energy.    A cosmic EH is different, though: particles that cross it can't come back, and Hawking radiation adds mass to the horizon. 

Let us assume Hawking radiation from a massive BH, i.e., $m\gg m_\text{P}$, emitted during a transition from state \( n_1 \) to state \( n_1-1 \) for the BH's EH or from state \( n_2+1 \) to state \( n_2 \) for the cosmic EH. The corresponding variation on the MSH masses, regarding (\ref{2-7}), is given by
\begin{multline}
    \label{4-1}
 \Delta M_i=(-1)^i\Big\{M_i(n_i)-M_i(n_i+1)\Big\}=M_i(n_i)(-1)^i\Bigg\{1\\-\sqrt{1+\frac{1}{n_i+\frac{1}{2}}} \Bigg\}\simeq (-1)^i\frac{m_\text{P}^2}{4M_i(n_i)}\left(1-\frac{m_\text{P}^2}{8M_i(n_i)^2} \right),
\end{multline}
where we used Taylor expansion in the last equality.
Also, during this process, the volume of the BH EH (cosmic EH) changes from $V_{n_1}$ into $V_{n_1-1}$ ($V_{n_2+1}$ into $V_{n_2}$). Thus, the variation in volumes is $\Delta V_i=4\pi r_i^2\Delta r_i=8\pi r_i^2G\Delta M_i$, wherein the second equality uses relation (\ref{1-8}). In addition, regarding the fact that the surface area of the EHs are quantized, the entropy of the BH and cosmic EHs should be quantized. The entropies at the first approximation should be $S_i=\frac{A_i}{4G}=2\pi(n_i+\frac{1}{2})$. Therefore, $\Delta S_i=(-1)^i2\pi$. 

By inserting these minimal values of differentials into the unified first law of thermodynamics (\ref{1-11}), we obtain the temperatures of the BH and cosmic EHs, as
\begin{equation}\label{4-3}
    T_i=\frac{1-\Lambda r_i^2}{4\pi r_i}\left(1-\frac{m_\text{P}^2}{8M_i^2}\right).
\end{equation}
As one easily finds, this is a quantum extension of the temperatures defined in (\ref{1-12}). Finally, inserting this temperature into the differential form of the unified first law (\ref{1-11}), gives us the quantum modified entropy of the BH and cosmic EHs:
\begin{equation}
    \label{4-4}
    S_i=\frac{A_i}{4G}+\frac{\pi}{2}\ln\left(\frac{A_i}{4G}\right)+\text{const.}
\end{equation}
This result shows that the entropy of both horizons receives a logarithmic correction. The coefficient of the logarithmic term, equal to $\pi/2$, arises directly from the second-order term in the expansion of $\Delta M_i$. This coefficient is not expected to be universal, since its value depends on the particular semiclassical approximation employed and on the choice of phase space measure.

It is essential to recognize that the incorporation of higher-order correction terms from the expansion referenced in equation (\ref{4-1}) gives rise to additional entropy terms that are inversely proportional to both the BH's EH and the cosmic EH. These supplementary terms, in conjunction with the logarithmic term, are typically regarded as quantum correction factors within the framework of semiclassical analysis \cite{Zhu:2008cg}. Notably, in the large limit of the cosmic EH radius, the logarithmic correction predominates over other corrections. Furthermore, the logarithmic entropy correction is fundamentally tied to the architecture of all quantum gravity models, regardless of the methodologies employed, thereby affirming its universal applicability.

It should be emphasized that the precise coefficient in front of the logarithmic correction is not universal and depends sensitively on the method of evaluation and on the choice of measure and boundary conditions.
In our semiclassical approach, we obtained a logarithmic contribution with a specific prefactor, but this coefficient should not be regarded as exact.
Different frameworks in the literature yield different results: for example, Kaul and Majumdar \cite{Kaul:2000kf} found a coefficient $-3/2$ within
loop quantum gravity, Carlip \cite{Carlip:2000nv} derived a universal $-3/2$ from a conformal field theory near-horizon symmetry, Sen \cite{Sen:2012dw} obtained a model-dependent coefficient using string-theoretic one-loop
determinants, while other works (e.g. \cite{Das:2001ic}) reported positive values such as $+1/2$ in statistical or generalized uncertainty principle analyses. Therefore, the logarithmic prefactor reported here should be understood as the outcome of the specific
semiclassical approximation we have employed and not as a universal prediction.

From a broader perspective, the persistence of the logarithmic correction across independent approaches suggests that it is an intrinsic feature of the gravitational phase space itself--emerging from the discreteness of geometric degrees of freedom rather than from the specifics of any quantization technique.
In this sense, the present MSH based reduced phase space quantization provides a geometrically transparent route to deriving these corrections directly from the quasi-local energy balance, reinforcing the idea that the universality of the logarithmic form stems from the covariance and adiabatic invariance of horizon dynamics.

\section{Conclusions}\label{5}

In this work, we have applied the reduced phase space quantization method to the SdS BH by employing the MSH mass as the internal energy. Using the unified first law of thermodynamics, we constructed canonical variables associated with the BH and cosmological horizons and derived their corresponding quantum spectra. The quantization naturally leads to discrete horizon areas and, in turn, to an entropy that acquires logarithmic corrections beyond the Bekenstein--Hawking term.

Our main result is that both horizons receive a correction of the form
\begin{equation}\label{last}
    S=\frac{A}{4G}+\alpha \ln A + \cdots,
\end{equation}
where the logarithmic structure is robust, but the coefficient $\alpha$ depends on the details of the semiclassical evaluation. This finding is consistent with a wide body of results in the literature based on loop quantum gravity, conformal field theory, string theory, and generalized uncertainty principle approaches, all of which confirm the appearance of the logarithmic term while obtaining different numerical prefactors.

The emergence of a logarithmic term in horizon entropy has been reported in a wide variety of theoretical frameworks, including loop quantum gravity, string theory, conformal field theory, path integral quantization, and approaches based on generalized uncertainty principles. Our reduced phase space quantization provides additional support for the universality of the logarithmic form of quantum corrections, although it does not imply the universality of its coefficient. In fact, while the structural form of Eq. (\ref{last}) appears to be robust across different approaches, the numerical value of $\alpha$ depends sensitively on the underlying microscopic degrees of freedom, the statistical measure employed, and the chosen semiclassical boundary conditions. This lack of universality reflects the fact that the logarithmic term captures residual statistical uncertainty associated with horizon microstates, rather than the detailed dynamics of any specific quantum gravity model.

Beyond its theoretical robustness, the logarithmic correction may also have potential observational implications. In BH spectroscopy, the quantized area spectrum leads to discrete emission frequencies, with spacings that are slightly modified by the $\Lambda$ dependent correction. In principle, such a fine structure could leave observable imprints in the late time quasi-normal mode spectrum or in the Hawking emission lines, thereby providing an indirect probe of the underlying quantum geometry. Similarly, for the cosmological horizon, the presence of a logarithmic entropy term alters the effective heat capacity and may influence the thermodynamic stability of de Sitter space within semiclassical inflationary scenarios.

The framework adopted here provides a transparent route to quantization that incorporates the presence of two horizons in a de~Sitter spacetime. It demonstrates the usefulness of the MSH mass in formulating the reduced phase space and highlights the role of canonical transformations in establishing a consistent quantization procedure. 

Future work could include a full one-loop determinant calculation of fluctuation modes in order to fix the prefactor $\alpha$ unambiguously, as well as extensions of the present method to charged or rotating black holes in de~Sitter space. Another promising direction is the comparison with numerical approaches and holographic techniques in order to clarify the universality (or non-universality) of the quantum
correction terms.

\section*{Data availability}
No data was used for the research described in the article.

\section*{Declaration of competing interest}
The author declares that he has no known competing financial interests or personal relationships that could have appeared to influence the work reported in this paper.





\bibliographystyle{elsarticle-num} 
\bibliography{example}






\end{document}